\documentclass[twocolumn]{aastex631}
\usepackage{amsmath}
\usepackage{fontspec}
\newfontfamily\russianfont{FreeSerif.ttf}

\def\kms{\,{\rm km}\,{\rm s}^{-1}}

\def\feh{\hbox{[Fe/H]}}

\newcommand{\teff}{T_{\rm eff}}
\newcommand{\rv}{{\rm RV}}

\def\vsini{(V \sin{i})}
\def\logg{\log{\rm (g)}}
\def\snr{\hbox{S/N}}

\def\drv{\vert \Delta {\rm RV} \vert}
\def\imp{f_{\rm imp}}

\begin{document}


\title{Search for double-line spectroscopic binaries in RAVE survey.}

\correspondingauthor{Mikhail Kovalev}
\email{mikhail.kovalev@ynao.ac.cn}

\author[0000-0002-9975-7833]{Mikhail Yu. Kovalev \russianfont{(Михаил~Юрьевич~Ковалёв)}}
\affiliation{International Centre of Supernovae (ICESUN), Yunnan Key Laboratory of Supernova Research, Yunnan Observatories, Chinese Academy of Sciences (CAS), Kunming 650216, China}


\begin{abstract}

I present a study for double-line spectroscopic binaries content in spectra from the RAVE survey, using a composite spectral model. Three complementing selections found 2813 composite spectra, belonging to 2605 unique targets. Also these results were compared with an automatic classification based on t-SNE map, which shows good agreement. Additionally, several spectra were identified to be affected by problems with the wavelength calibration.  


\end{abstract}



\section{Introduction} \label{sec:intro}

The RAdial Velocity Experiment (RAVE) survey had its final 6$^{th}$ data release (DR6) six years ago \citep{rave6}, providing more than half million spectra for stars in the southern sky, observed from 2003-4 till 2013-3. The main focus of the survey was to measure accurate radial velocities ($\rv$), which allow for a detailed study of kinematic properties of the Galaxy \citep{xenia}. Spectroscopic binaries were not the main interest for RAVE, but \cite{rave_sb2} looked for double-line spectroscopic binary (SB2) candidates and found 123 detections, exploring the second data release with 25850 spectra. In a follow up study, \cite{rave_sb1} identified 1333 single-line spectroscopic binaries (SB1) candidates. Later, \cite{sb1rave2} found 27716 SB1 candidates, by combining RAVE radial velocities with additional information from Gaia DR2 \citep{gdr2}. Moreover, an automatic classification with the help of an embedding algorithm by \cite{ravelle} reports 3123 spectra having ``binary" as a primary classification flag in RAVE DR6.
\par
Recently, \cite{m11} searched for SB2 candidates in the open cluster M11 using near infrared spectra from the Gaia-ESO survey \citep{ges}. All spectra were fitted by a composite spectral model and 24 SB2 candidates were found. 
This model was further adapted to fit LAMOST-MRS \citep{mrs} spectra in \cite{cat22}, which also presented a new method for SB2 candidate detection using combination of { projected rotational velocity} $\vsini$ values from fits by single-star and binary spectral models. It found 2460 candidates in time domain subsurveys of LAMOST DR10, where 1410 were new discoveries. The follow up study \cite{cat23} applied this method for all spectra from LAMOST DR11 and detected 12426 (8105 new) SB2 candidates. \cite{rot5oc} used the same method to find 60 SB2 candidates (42 new) in five open clusters observed by Gaia-ESO. Recently, \cite{guo665} confirmed 665 candidates by providing orbital solutions for each.

\par

In this work I apply methods from the previously mentioned papers to the spectra from RAVE DR6 to explore its SB2 content. This survey wasn't intended for binary detection, so the number of SB2 candidates is a bit smaller than those found in in LAMOST. Information on SB2 candidates helps to flag and exclude them in advance, to prevent possible biases in spectroscopic parameters \citep{kareem1} based a single star assumption.  Additionally, $\rv$ measured in 2003-2013 are perfect for verification of spectroscopic orbits computed using later surveys. Slow, long period changes in systemic velocity can also be identified, which helps to identify possible hierarchical triple systems candidates. 


\section{Observations (DR6 SPECTRA)}

I downloaded all available 518302 spectra\footnote{each of them have unique identification ROI} from \url{www.rave-survey.org/}. They cover { a narrow spectral window around Ca II triplet} $\lambda=8420:8780$~\AA~ and have resolution $\sim7500$. They are already normalised and shifted to the restframe, { according to heliocentric radial velocity (HRV), given in RAVE DR6}. The mean signal to noise ratio ($\snr$) is 42 pix$^{-1}$, ranging from 1 to 764 pix$^{-1}$. In total 451747 unique targets were observed, where 404386 have only one exposure, while the maximal number of exposures reaches 13 for one star.   

\section{Methods and results} 
\label{sec:methods} 

\subsection{Spectroscopic analysis}
\label{sec:specfit}
All spectra were fitted by the single-star and binary models, adopted from \cite{m11} originally generated for Gaia-ESO spectra, { see Appendix~\ref{sec:mod}}. The RAVE spectra have slightly different coverage and sampling, therefore models were down-sampled and clipped. Additionally I keep only models with $|\feh|\leq0.3$ dex in the grid (12936 in total), because metallicity is not very important for SB2 identification, and it significantly speeds up the computations (each spectrum takes $\sim4$ s to process). The resulting grid covers $\teff=5000:15000$ K with step 500 K, $\logg=3:5$ dex with step 0.2 dex, $\feh=-0.3:0.3$ dex with step 0.1 dex. { The stellar rotation has only two values for each combination of other parameters: almost no rotation $\vsini=1~\kms$ and a value randomly generated from uniform distribution $\vsini=\mathcal{U}(1,380)~\kms$. This choice allows the grid to be relatively small, while the not-rotating case is not affected by the possible interpolation errors, similarly to \cite{cat23}}. 
The single-star model is linearly interpolated from this grid, Doppler-shifted and multiplied by coefficient $C$, which takes in to account the possible difference in normalization with observation. 
The  binary model spectrum is computed as a sum of the two Doppler-shifted  single-star spectral models ${f}_{\lambda,i}$, scaled according to the difference in luminosity, which is a function of the $\teff$ and a proxy for stellar size ratio - combination of Planck function ${B_\lambda(\teff)}$ and coefficient $k$: 

\begin{align}
    {f(\rv_0)}_{{\rm single}}=C{f(\rv_0)},\\
    {f(\rv_1,\rv_2)}_{{\rm binary}}=C\frac{{f_2(\rv_2)} + k_\lambda {f_1(\rv_1)}}{1+k_\lambda},\\
    k_\lambda=k \frac{B_\lambda(\teff{_{,1}})}{B_\lambda(\teff{_{,2}})}.
	\label{eq:bolzman}
\end{align}

The resulting spectrum is compared with the observed one using \texttt{scipy.optimise.curve\_fit} function, which provides the optimal spectral parameters and radial velocities (RV) for each component plus coefficients $C,k$. We keep the metallicity equal for both components during binary fit. 
In total I have 6 free parameters for a single star fit and 11 for a binary fit. I estimate the goodness of the fit parameter by reduced $\chi^2$.
Also I compute the improvement factor ($\imp$) using Equation~\ref{eqn:f_imp} similar to \cite{bardy2018}. This improvement factor estimates the absolute value difference between two fits and weights it by the difference between the two solutions.

\begin{align}
\label{eqn:f_imp}
f_{{\rm imp}}=\frac{\sum_{\lambda=\lambda_{min}}^{\lambda=\lambda_{max}}\left[ \left(\left|{f}_{\lambda,{\rm single}}-{f}_{\lambda}\right|-\left|{f}_{\lambda,{\rm binary}}-{f}_{\lambda}\right|\right)/{\sigma}_{\lambda}\right] }{\sum_{\lambda=\lambda_{min}}^{\lambda=\lambda_{max}}\left[ \left|{f}_{\lambda,{\rm single}}-{f}_{\lambda,{\rm binary}}\right|/{\sigma}_{\lambda}\right] },
\end{align}
where ${f}_{\lambda}$ and ${\sigma}_{\lambda}$ are the observed flux and corresponding uncertainty, ${f}_{\lambda,{\rm single}}$ and ${f}_{\lambda,{\rm binary}}$ are the best-fit single-star and binary model spectra, and the sum is over all wavelength pixels.

\subsection{Selection of SB2 candidates.}

\begin{figure}
    \centering
    \includegraphics[width=0.95\linewidth]{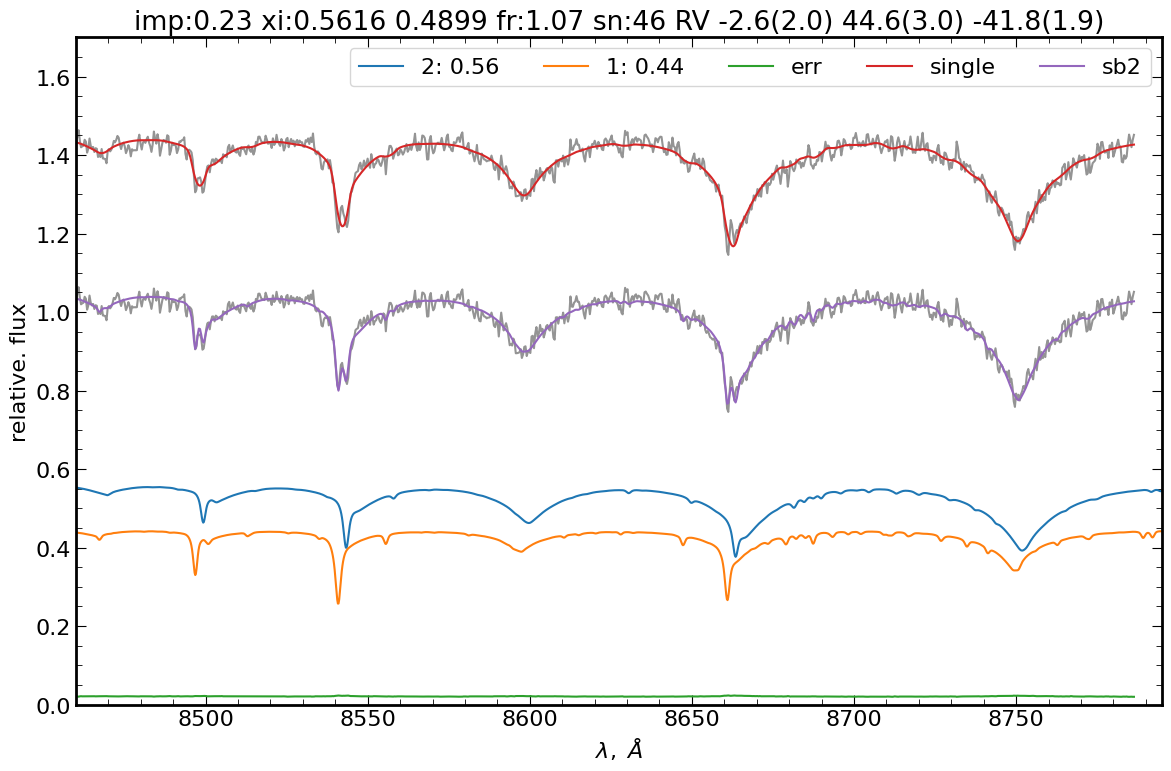}\\
    \includegraphics[width=0.95\linewidth]{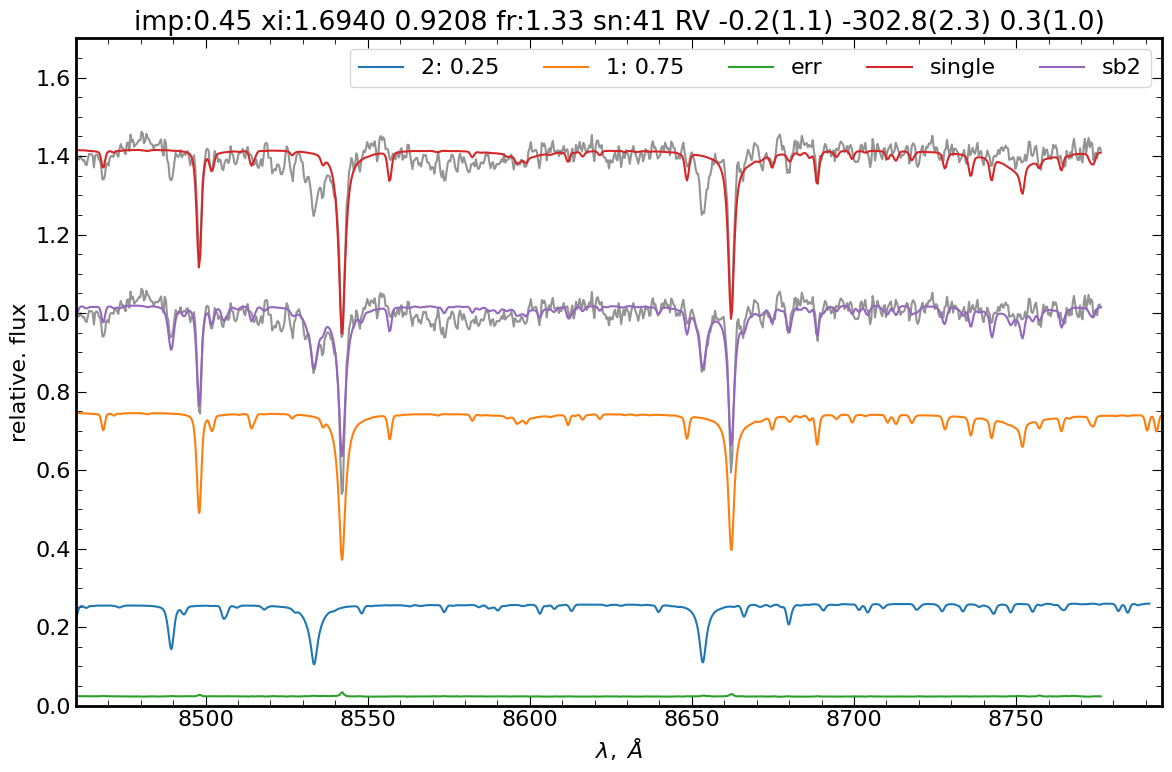}\\
    \caption{Examples of RAVE spectra and fits for J054721.6-410945 (top panel - $\vsini$ selection) and J120454.9-780443 (bottom panel - selection by primary). The titles present $\imp, \chi_{\rm single, binary}^2, k, \snr$ and radial velocities $\rv_{0,2,1}$ with errors. Note that since RAVE spectra have wavelength calibration to rest-frame all these $\rv$ need to be corrected by HRV given in RAVE DR6. Relative contributions of two components are shown in the legend.}
    \label{fig:exampl}
\end{figure}

My first SB2 selection is based on criteria introduced in \cite{cat22}: if we make a plot of $\vsini_0$ vs $\vsini_1+\vsini_2$\footnote{{ Hereafter I use subscript ``0" for parameters related to single-star model, and ``1,2" for parameters related to components of binary model. }} there is a forbidden region for the single star results: $\vsini_0<\vsini_1+\vsini_2$, assuming that the binary and single star models fit the observed spectrum reasonably well. In this regime the single-star model increases the value of $\vsini_0$, so the broadened model covers both components of SB2, see the top panel of Figure~\ref{fig:exampl}. As a consequence, we can find a correlation between $\vsini_0$ and the absolute difference of radial velocities in SB2. I call it $\vsini$ selection.
\par
Unfortunately, this selection is usually unable to detect some SB2 candidates, when the single-star model fits only one component (usually the primary, where it outshines the secondary), completely ignoring the secondary, while the binary spectral model still produces a good fit for both components, see the bottom panel of Figure~\ref{fig:exampl}. To keep these detections, \cite{cat23} introduced another selection, which mostly focuses on large $\imp$ and $\chi^2_{single}/\chi^2_{binary}$. I call it primary selection.
Details for these selection criteria are provided in Table~\ref{tab:cuts}.
\par
Thanks to the automatic morphology classification flags provided in RAVE DR6 we can make another simple selection. Each spectrum has up to three flags ($P, S, T$ - primary, secondary and ternary)\footnote{RAVE DR6 uses a simplified version of the classification scheme introduced in \cite{ravelle}, providing only the first three classification flags instead of the original 20, which are ordered according to decreasing weight. } accompanied by weights $w_P+w_S+w_T\leq1$, which have different codes i.e 'b'-binary. 3123 spectra have a binary primary flag with $P=$`b', so it is reasonable to expect that they have a good fit with the binary model. 
\par
{ In this study, I prioritize purity of the sample over the completeness, therefore plots for all selected SB2 candidates were visually checked to exclude false positives, similar to \cite{cat23}. The selected candidate must show double-line structure: at least two lines of Ca II triplet are distorted in the same way, or a clear splitting of weaker lines (i.e Fe lines at 8514, 8688~\AA), or clear combination of the hot component (with Paschen lines) and cool component (prominent Ca II triplet) is visible. }
\par
\begin{figure*}
    \includegraphics[width=0.5\textwidth]{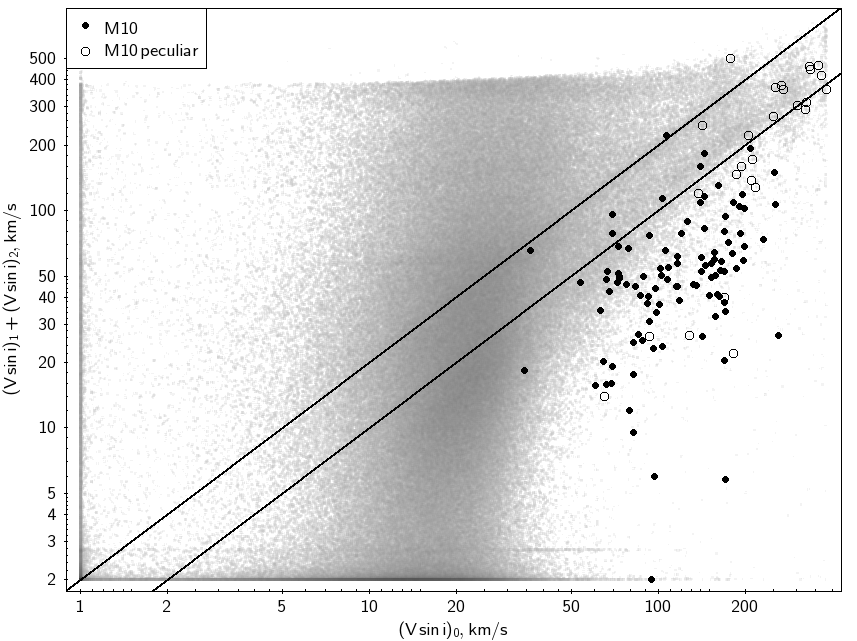}
    \includegraphics[width=0.5\textwidth]{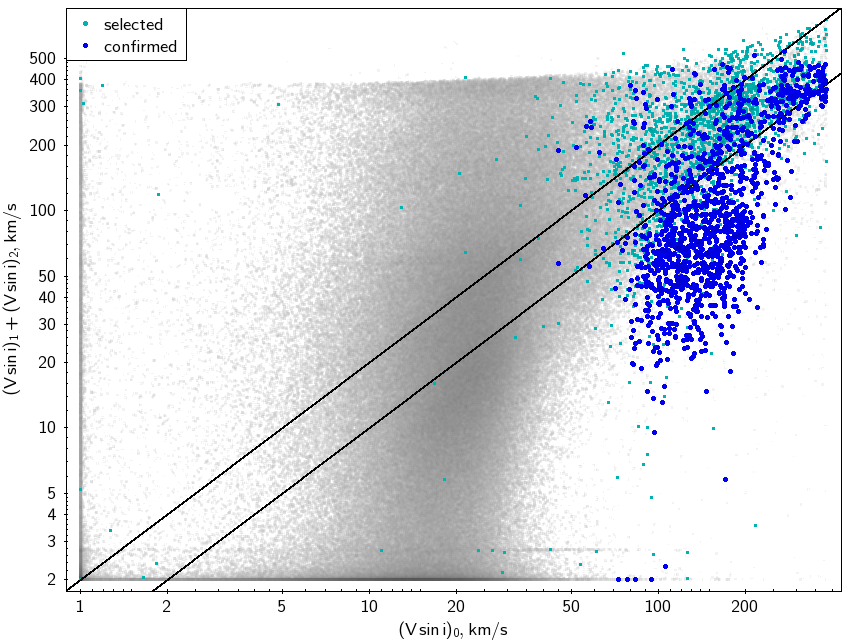}\\
    \includegraphics[width=0.5\textwidth]{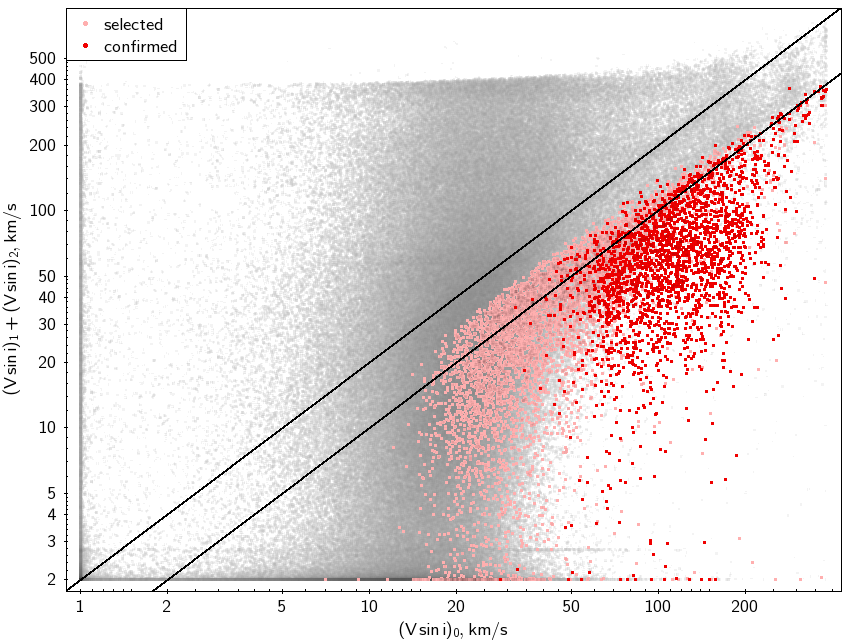}
    \includegraphics[width=0.5\textwidth]{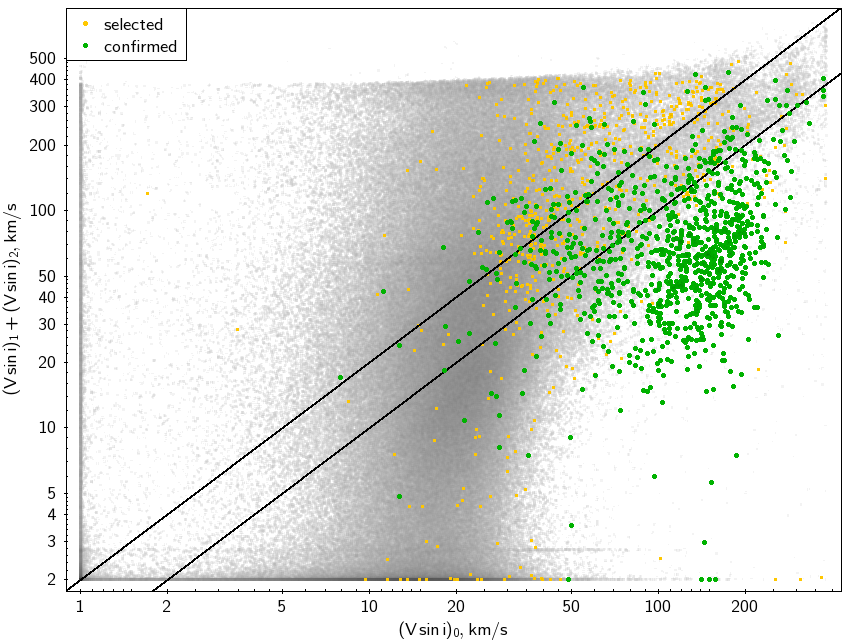}
    \caption{ Different selections of SB2 candidates: top left - \protect\cite{rave_sb2}, top right - automatic classification with P='b', bottom left - $\vsini$ selection, bottom right - primary selection. Diagonal solid lines show functions $\vsini_1+\vsini_2=\vsini_0$ and $\vsini_1+\vsini_2=2\vsini_0$.} 
    \label{fig:rot}
\end{figure*}

Using all three selections results in 2813 spectra after visual inspection, belonging to 2605 unique targets, with some of them chosen by all three. In Figure~\ref{fig:rot}, projected rotational velocity $\vsini$ is plotted for different selections.
Additionally during visual inspection I identified 13 possible SB3 systems:
\begin{itemize}
    \item J113828.2-301729, ROI 20040627\_1138m31\_074   
    \item J140414.5-140845, ROI 20040704\_1356m14\_115
    \item J063128.3-633207, ROI 20050129\_0611m63\_120
    \item J161628.1-065844, ROI 20050418\_1621m07\_054
    \item J101229.3-194927, ROI 20050420\_1010m20\_082
    \item J142625.4-104450, ROI 20060527\_1430m10\_083
    \item J204009.8-325939, ROI 20061003\_2046m31\_012
    \item J042510.1-510546, ROI 20071017\_0435m50\_032
    \item J141254.5-214205, ROI 20080524\_1414m20\_013
    \item J113520.3-234959, ROI 20090421\_1135m25\_077
    \item J102916.7-370012, ROI 20100120\_1036m37\_048
    \item J130736.0-141443, ROI 20110323\_1304m14\_114
    \item J224257.2-051839, ROI 20120814\_2247m04\_058
\end{itemize}
with spectra showing hints for three components, { where two components of binary model were not enough to capture all spectral features}. Note that \cite{rave_sb2} already reported J113828.2-301729 as a triple candidate.   

\begin{table}
    \centering
    \caption{Quality cuts and selection criteria.}
    \begin{tabular}{l}
\hline
\hline
good spectra:\\
$\snr>15$ pix$^{-1}$\\
$P\neq $'w' (not problematic spectrum)\\
$P\neq $'c'  (not problematic continuum)\\
\hline
good fits:\\
$\chi^2_{\rm single, binary}<20$  \\
$\sigma\rv_1$+$\sigma\rv_2 < 20$ $\kms$\\
\hline
$\vsini$ based SB2 selection (7719 spectra, 2304 after check):\\
good fits and good spectra\\
$\imp \geq0.1$  \\
$\vsini_1+\vsini_2-{\rm max} \left[\sigma\vsini_1,\sigma\vsini_2\right]<\vsini_0$\\
${\rm max} \left[\sigma\vsini_1,\sigma\vsini_2 \right]<0.5\vsini_0$\\
\hline
SB2 selection by primary (1668 spectra, 885 after check):\\
good fits and good spectra\\
$\imp>$ 0.20\\
$\chi^2_{\rm  binary}<10$  \\
$\chi^2_{\rm single}/\chi^2_{\rm binary}>1.2$\\
\hline
DR6 classification (3123 spectra, 1218 after check):\\
$P= $'b' (binary code)\\
\hline

    \end{tabular}
    \label{tab:cuts}
\end{table}

\subsection{Verification}

\subsubsection{Comparison with \cite{rave_sb2}}

\begin{figure}
    \centering
    \includegraphics[width=0.95\linewidth]{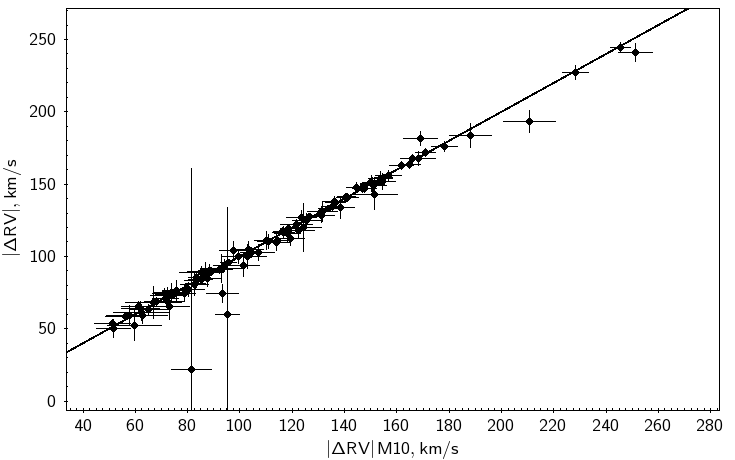}
    \caption{$\drv$ comparison from this work and values from \protect\cite{rave_sb2}.}
    \label{fig:drv}
\end{figure}

\cite{rave_sb2} have two tables with SB2 detections: one with 102 entries (101 have spectra in RAVE DR6\footnote{\dataset[plots for them and some additional targets can be found here]{https://mkuasbest.github.io/webpage/rave6m10.htm}}) with measured $\drv$ and 27 ``peculiar" candidates without $\drv$ measurement (26 have spectra in RAVE DR6). Unfortunately one candidate from each of these tables is missing in RAVE DR6. I compare $\drv$ from my fit and \cite{rave_sb2} in Figure~\ref{fig:drv}, which shows good agreement for most spectra, with four spectra having smaller $\drv$ in my fits.  
The primary selection has 51 + 3 matches with these two tables. The $\vsini$ selection has 90+13 matches with these two tables, while flag `b' selection has 43 + 19. All three selections choose 92+23 SB2 candidates, complementing each other, although some SB2 candidates from \cite{rave_sb2} were not selected by any. 

\subsubsection{t-SNE map}

\begin{figure*}
    \includegraphics[width=0.5\textwidth]{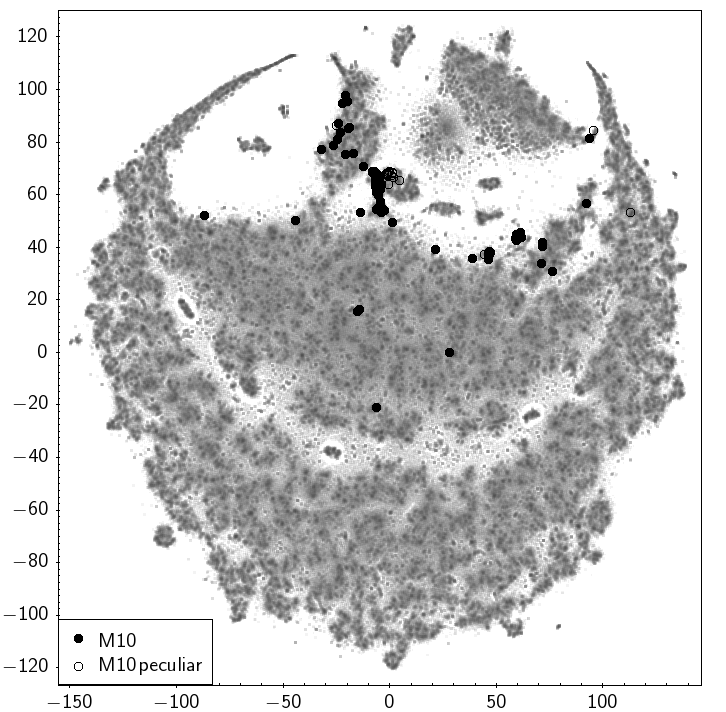}
    \includegraphics[width=0.5\textwidth]{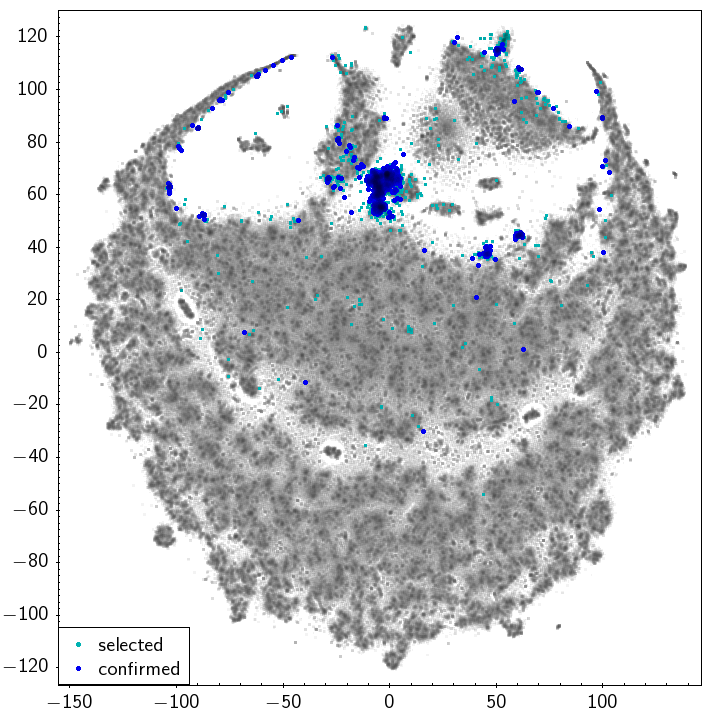}\\
    \includegraphics[width=0.5\textwidth]{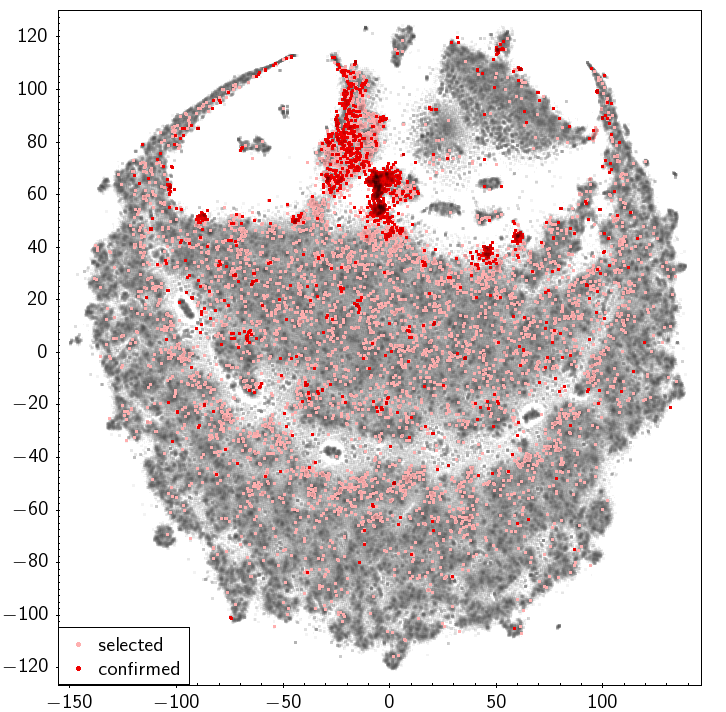}
    \includegraphics[width=0.5\textwidth]{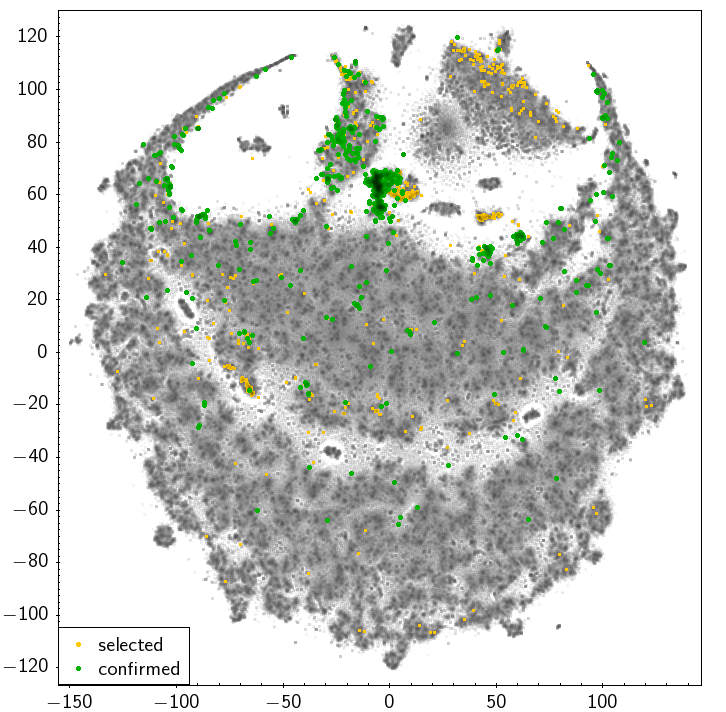}
    \caption{t-SNE map and different selections of SB2 candidates: top left - \protect\cite{rave_sb2}, top right - automatic classification with P='b', bottom left - $\vsini$ selection, bottom right - primary selection.  } 
    \label{fig:tsne}
\end{figure*}

Since a two dimensional (2D) map for the automatic classification of all RAVE DR6 spectra is not provided in \cite{rave6}, I computed one using t-distributed stochastic neighbor embedding (t-SNE \cite{tsne}), similar to \cite{traven_tsne}, who applied it previously to select SB2 candidates in spectra from Galactic Archaeology with HERMES (GALAH) survey \citep{galah2015}. { This technique allows one to explore huge datasets using simple visualization in form of a 2D or 3D map, where datapoints sharing similar features are grouped together in multiple clusters. A datapoint with some unique, distinct features will be placed by t-SNE as an outlier. Flux-normalized and restframe corrected RAVE DR6 spectra are well suited for the t-SNE technique, although some preparation of the dataset is necessary. Lower and upper wavelength edges are slightly variable across the spectra, plus edges can suffer from normalization issues. The number of datapoints in spectrum is not consistent among the dataset. Thus I selected only 800 datapoints in each spectrum ([110:910] pixels). This final dataset was fed to {\sc sklearn.manifold.TSNE} in default it's configuration. } The resulting map is shown in Figure~\ref{fig:tsne}, with highlighted selections similar to Figure~\ref{fig:rot}. It is clear that many SB2 candidates group into several clusters, most notably those around t-SNE coordinates 0,60 (especially $P=$`b' selection, which is not surprising, as this method is similar to t-SNE), although some of them are located even at the bottom of the map, so pure automatic classification can miss them.   
{ I show this map for illustrative and verification purposes, it was not used to add/reject any SB2 candidate.}

\subsubsection{SX Oph and HD 20784}

\begin{figure}
    \centering
    \includegraphics[width=0.95\linewidth]{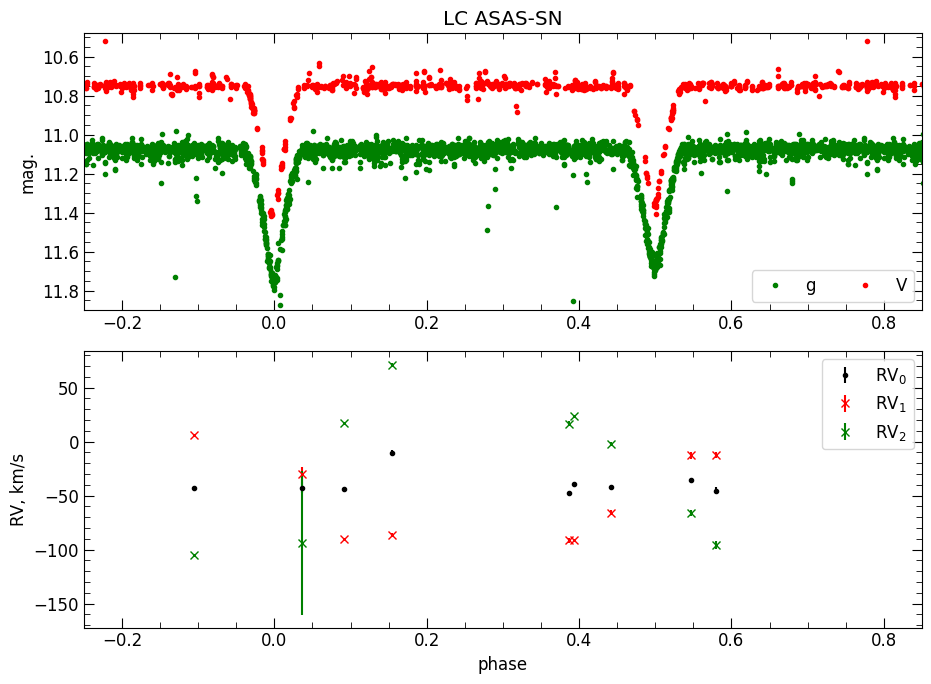}\\
    \includegraphics[width=0.95\linewidth]{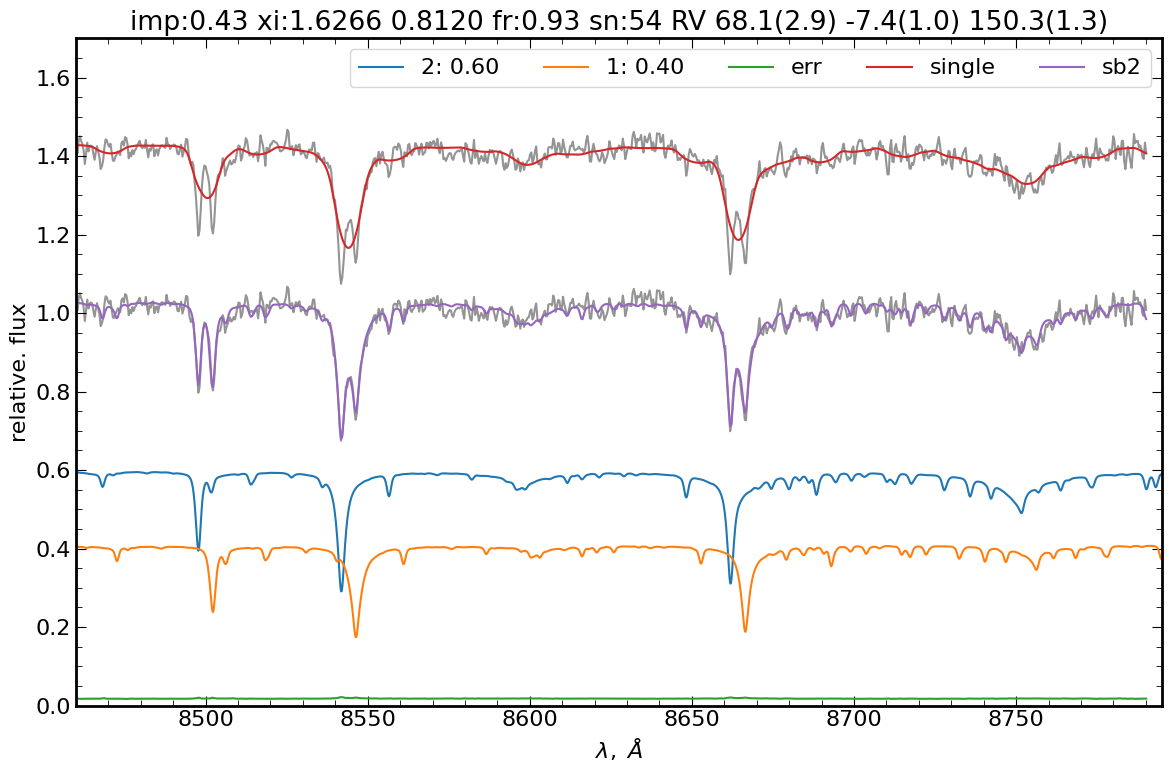}
    \caption{Light curve and radial velocities for detached eclipsing binary SX Oph (top and middle panels). Example of fit for epoch with maximal $\drv$ (bottom panel).}
    \label{fig:sxopi}
\end{figure}

Detached eclipsing binary SX Oph has 9 spectra in RAVE DR6 under id=J161755.5-063952. I show my $\rv$ and light curve (LC) from ASAS-SN \citep{asassnv} for this system in Figure~\ref{fig:sxopi}, where the phase is computed with ephemeris: $t_0=2456704.5828 +4.1266E$  from \cite{rowan_asassin}. All $\rv$ were added with HRV values from \cite{rave6}, because they were measured from spectra with the wavelength corrected to rest frame using these values. Additionally I sorted $\rv$ measurements from the binary model, so the brighter component will have index 1, and the dimmer will have index 2. The single-star model $\rv_0$ shows mostly only small changes with time, while $\rv_{1,2}$ shows changes consistent with LC (the primary component approaches us after the deeper eclipse, while the secondary recedes and inverted behavior is seen after the shallower eclipse). One measurement taken at the end of the primary eclipse was poorly fitted by the binary model: $\rv_1~\rv_0$, while $\rv_2$ is significantly off. The phase coverage is not optimal to get a good orbital solution, although the amplitude of $\rv$ changes suggest that primary component is more massive with $q\sim0.5$. 
Four spectra of SX Oph have been selected with $P=$'b' (remaining five spectra have $S=$'b'), seven spectra were selected by $\vsini$ and seven by primary.  
\par
\begin{figure}
    \centering
    \includegraphics[width=0.95\linewidth]{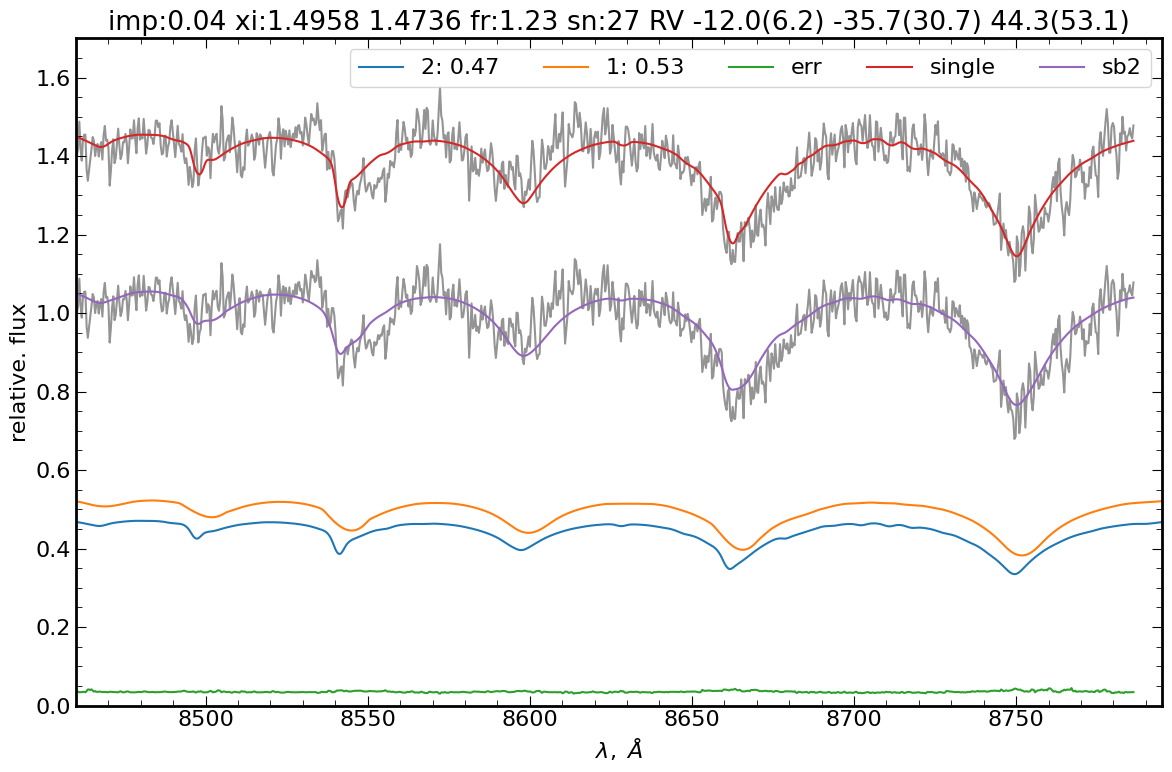}\\
    \includegraphics[width=0.95\linewidth]{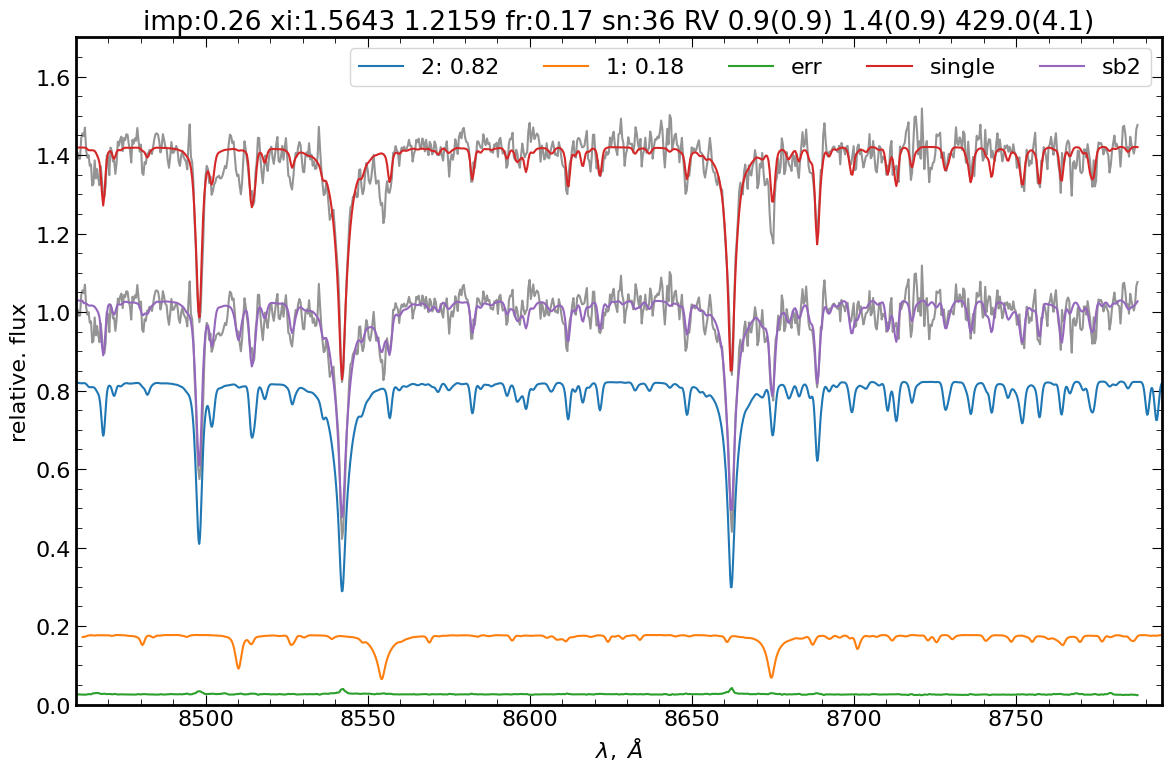}\\
    \caption{RAVE spectra and fits: HD 20784 (top panel) and J132346.4-675653 (bottom panel).}
    \label{fig:hd20784}
\end{figure}

HD 20784 is a SB2 binary consisting two hot components: $\teff\sim9500,10400$ K with orbit derived in \cite{galaxies14020027} using high resolution spectra from South African Large Telescope (SALT). It has one spectrum in RAVE DR6 classified as a hot star, with radial velocity HRV=$87\pm34~\kms$, which is inconsistent with SB2 orbit's computed values $\rv_{1,2}=0.3,-40.2~\kms$. My fit provides $\rv_{0,1,2}=74.9\pm6.2, 51.2\pm30.7,131.2\pm53.1~\kms$, which is also very discrepant, see Figure~\ref{fig:hd20784} top panel. This can indicate a significantly lower quality of RAVE spectra for analysis of hot stars, but also a hint of a possible long period motion of barycenter in HD 20784 (SALT observed it 16 years later than RAVE).

\subsubsection{chance alignment}

{ The relatively large fiber diameter of 6.7\arcsec used in RAVE, means that a significant number of detected SB2 candidates can be in fact just a chance alignments.  
For example,} one spectrum with id=J132346.4-675653 was selected by the primary method, but it has $P=$'n' - normal stars with $w_P=1$, which is surprising as there are two components clearly visible in the spectrum (Figure~\ref{fig:hd20784} bottom panel) with $\rv_{1,2}=-13.5\pm0.9,~414.1\pm4.1~\kms$ (these values were added with HRV). According to \cite{gaia3}, here we have two stars of comparable brightness $G=11.18,~12.59$ mag, but different parallaxes: $\varpi=1.107, 0.133$ mas and proper motions: $\mu_\alpha \cos{\delta}=-2.70,-7.79$ mas/yr, $\mu_\delta=-7.87,-1.71$ mas/yr, therefore it is { definitely a} chance alignment pair observed by the same fiber, { not a SB2 candidate}. The Gaia DR3 $\rv=-12.57\pm0.23,~408.71\pm1.05~\kms$ from Radial Velocity Spectrometer (RVS \cite{grvs}) are a bit higher than my estimates. It is worth noting, that \cite{sb2jack_2019AN....340..386J} mentioned it previously as a SB candidate, based on difference of radial velocities in RAVE and Gaia.
\par
{ Unfortunately, not all chance alignments can be easily identified using Gaia data, therefore one should keep in mind their possible presence in the SB2 candidates catalog.}

\subsubsection{problems with wavelength calibration}
\cite{problem_rv5} reported that 707 spectra from RAVE DR5 have problems with wavelength calibration, which resulted in a shift of $\Delta\rv=105~\kms$, when compared with Gaia DR2. I checked their ROI and confirm that they were excluded from DR6. 
\par
\begin{figure}
    \centering
    \includegraphics[width=0.95\linewidth]{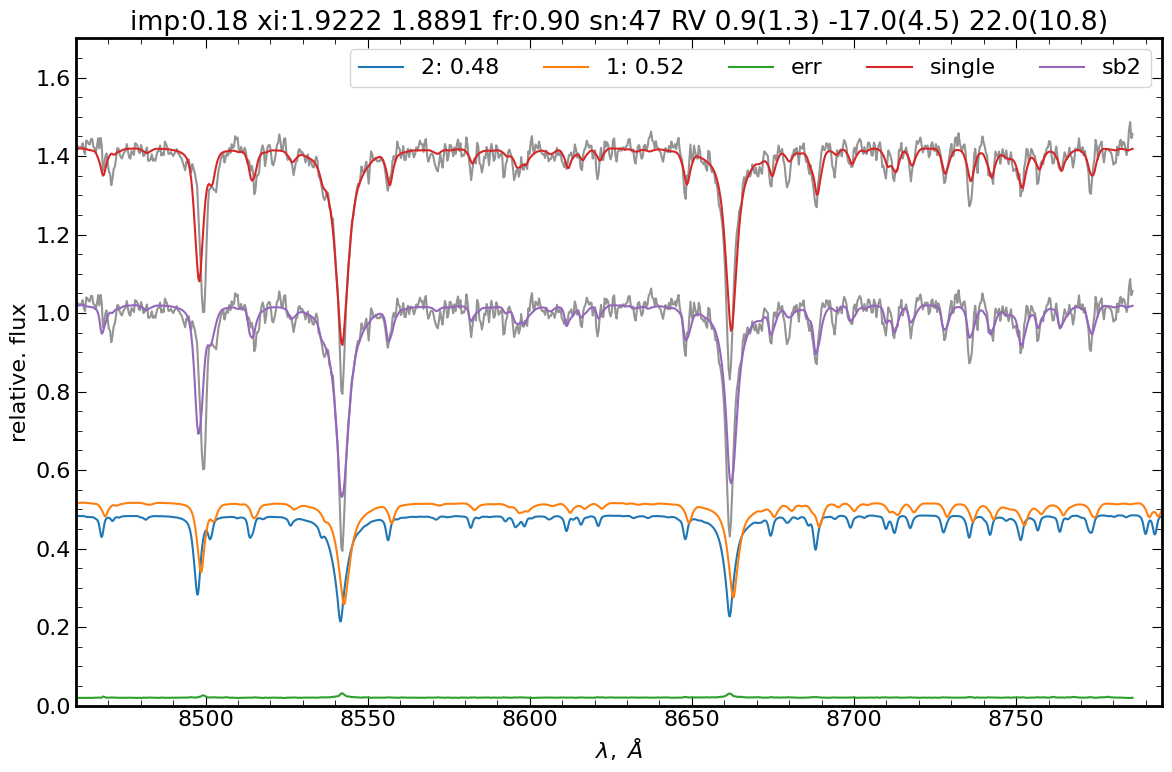}\\
    \includegraphics[width=0.95\linewidth]{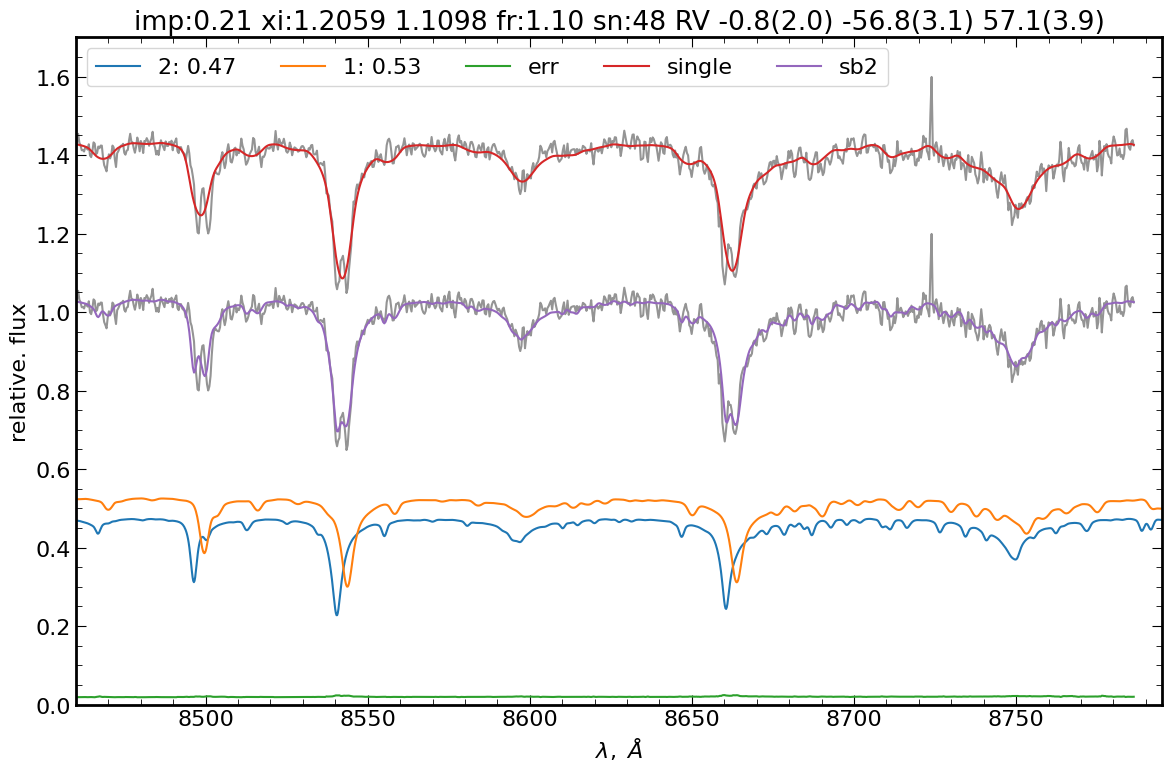}
    \caption{RAVE data and fits for two spectra with wavelength calibration problems: normal star J051136.3-293551 (top panel) and SB2 candidate J134949.3-621355 (bottom panel).}
    \label{fig:wvl}
\end{figure}

During the visual inspection of plots for the $\vsini$ selection another problem with the wavelength calibration was identified: in some spectra the ``blue" part showed a red shift, while the ``red" part is shifted to blue, which is clearly visible for Ca II lines at $\lambda=8498,8663$~\AA~ while the central line of triplet is not shifted, see Figure~\ref{fig:wvl}. The top panel shows a normal single star, and the bottom panel shows the SB2 candidate\footnote{It worth to note that other two spectra of this target were not affected by this problem and both were selected as SB2.} both affected by this problem. This SB2 candidate is included in \cite{rave_sb2} ``peculiar" spectra, where the issue with the wavelength calibration was also reported. Similarly to the LAMOST-MRS analysis in \cite{cat22}, the binary model fits such problematic spectra a bit better than the single-star model, so they get selected as SB2 candidates. Fortunately these spectra are clustered in the t-SNE map in a small area around coordinates 97,83 (238 spectra in total) and can be easily found.

\section{Discussion}
In order to { test the } quality of the $\rv$ measurements and their applicability for SB2 orbit { determination/verification, I checked} several literature sources of { known binary systems}.
\par
\begin{figure}
    \centering
    \includegraphics[width=\columnwidth]{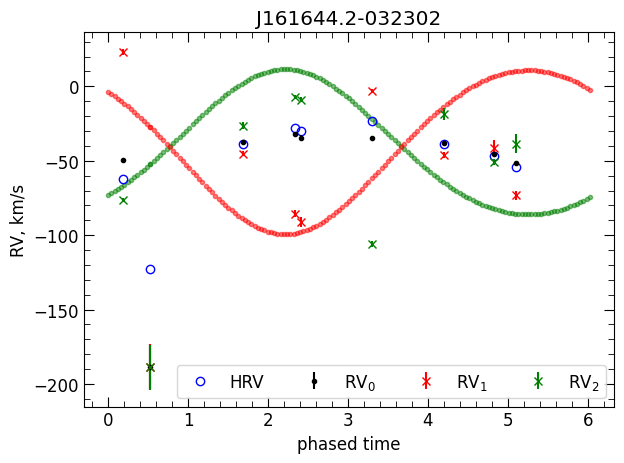}\\
    \includegraphics[width=\columnwidth]{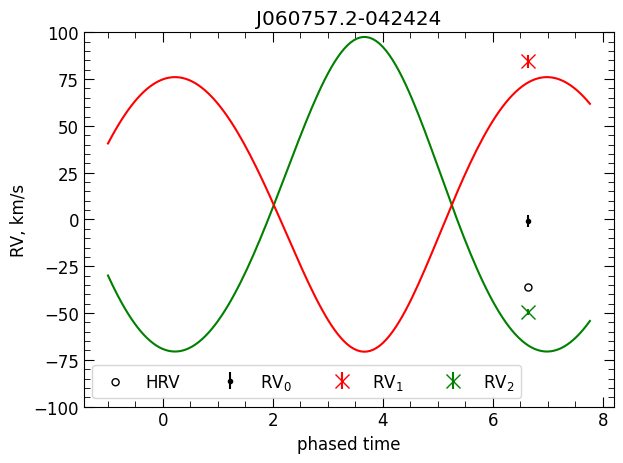}\\
    \includegraphics[width=\columnwidth]{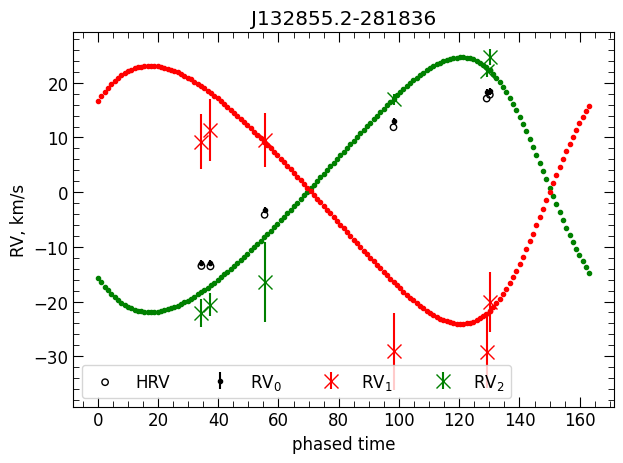}
    \caption{Orbital solutions from literature and $\rv$ measurements for J161644.2-032302 (top panel), J060757.2-042424 (middle panel) and J132855.2-281836 (bottom panel).}
    \label{fig:orb}
\end{figure}

Gaia DR3 provides Non Single Star catalog (NSS) which includes many binary orbits based on astrometric, spectroscopic and photometric observations carried out by that space observatory. My selection of 2605 SB2 candidates have: 

\begin{itemize}
    \item 7 matches with ``AstroSpectroSB1" 
    \item 36 matches with ``EclipsingBinary"
    \item one match with ``EclipsingSpectro" 
    \item 32 matches with ``Orbital" 
    \item one match with ``OrbitalTargetedSearch"
    \item 104 matches with ``SB1" 
    \item 220 matches with ``SB2"
    \item 33 matches with ``SB2C"
    \item 53 matches with ``Acceleration7"
    \item 24 matches with ``Acceleration9"
    \item 3 matches with ``FirstDegreeTrendSB1"
    \item 2 matches with ``SecondDegreeTrendSB1"
    \item one match with ``VIMF"
\end{itemize}
tables, where matches with the SB2 table are particularly interesting, because the Gaia RVS instrument observes the same spectral range. For example, J161644.2-032302 with nine spectra is shown on the top panel of Figure~\ref{fig:orb}. It is clear that one $\rv$ measurement is very bad ($\rv_{1,2}\sim-190~\kms$) which was derived from a very noisy spectrum. The other $\rv$s qualitatively agree with SB2 orbit, although the assignment of components is sometimes wrong. Also the periastron passage time should be adjusted, similarly to \cite{sb3japan}, since Gaia DR3 SB2 orbit was based on significantly later measurements than RAVE DR6.      
\par
There is only one match with the SB2 orbit catalog from \cite{guo665}: J060757.2-042424 with one spectrum available in RAVE DR6, selected by primary and flag `b'. RV measurements are shown together with orbital solution in Figure~\ref{fig:orb} middle panel. The $\rv_{1,2}$ are larger than orbital solution by 10 $\kms$, while $\drv$ is consistent. This indicates the possibility of the SB2 being part of a hierarchical triple, although it can be also a phase shift like for J161644.2-032302. 
\par
My selection has 9 matches with SB9 \citep{sb9}. Interesting, that one SB2 system from this catalog J132855.2-281836 has 6 spectra in RAVE DR6, with two of them first selected by $\vsini$, but were rejected during visual inspection as unreliable detections, { so it is absent in the final list of 2605 SB2 candidates}. I show $\rv$ measurements and orbital solution in the bottom panel of Figure~\ref{fig:orb}. Here I sorted the $\rv$ so that the brighter component will have index 2 and the fainter component have index 1. The single star measurements are very close to the HRV values from RAVE DR6, and they both follow $\rv_2$, which shows good agreement with the orbital solution. The fainter component $\rv_1$ generally is more uncertain, but still shows an acceptable match with orbit, taking errors into account. Therefore my radial velocity measurements can be useful even for the small $\drv$ regime, however one should use them with care for the purpose of orbit determination or verification. 
 
\par
In principle, one can utilize SB2 selections rejected during visual inspection, if multiple spectra are available and $\vsini_0$ shows variability with time. It is similar to \cite{j115}, where a wide twin system with period $P\sim206$ days, was discovered based on LAMOST-MRS spectra. In this case there will be more detections. 

\par 
{ If I combine information from known the SB1/SB2 catalogs (SB9, NSS, \cite{rave_sb2,rave_sb1,kounkel,cat23}) together, there are 601 match with my SB2 table of 2605 candidates. Therefore around $\sim2000$ SB2 candidates are new detections, although possibly my list of known SB1/SB2s could be incomplete. }

I present a catalog with results on RAVE DR6 spectra analysis in Table~\ref{tab:cat} \footnote{\dataset[currently available via link with code: Udgx ]{https://mail.cstnet.cn/coremail/XT5/jsp/download.jsp?share_link=F93EF5BD9EE54DC0BE3894E0A12DFA21&uid=mikhail.kovalev\%40ynao.ac.cn} } , which includes some information from RAVE DR6 (ID, number of repeats,classification flags $P,S,T$, RA, DEC, ROI, HRV, $\snr$ and mean MJD for each spectrum calculated from observation start/end times), fitting parameters from the single-star and binary models, t-SNE coordinates and SB2 index, if it was confirmed during visual inspection. The $\rv$ values should be added with HRV values, if one wants to use them for further studies, because they were measured from spectra with the wavelength corrected to restframe using these values. The spectral parameters ($\teff,\logg,\feh,\vsini$) and their errors represent best fit parameters of the ``flexible template", which should not be used as exact measurements, { but like rough initial estimations. They play the same role } as parameters of the synthetic template often used for RV determination using a cross-correlation function.

\setlength{\tabcolsep}{9pt}
\begin{table}
\centering
\caption{Catalog for RAVE DR6. This table is available in its entirety in machine readable form}
\begin{tabular}{ccc}
\hline
Parameter & name & unit\\
  RA & ra\_input & degree\\
  DEC & dec\_input & degree\\
  N& {n\_repeats} &\\
  {ID} & ID\\
  {HRV} & HRV&$\kms$\\
  {S/N} & S/N& pix$^{-1}$\\
  $P$ & P&\\
  $S$ & S&\\
  $T$ & T &\\
  {ROI} &roi&\\
  {t-SNE x} & tsnex&\\
  {t-SNE y} &tsney&\\
  {$\rv_0$} & rv0& $\kms$\\
  {$\sigma\rv_0$} & erv0&$\kms$\\
  {$\rv_1$} & rv1&$\kms$\\
  $\sigma\rv_1$& erv1&$\kms$\\
  {$\rv_2$} &rv2&$\kms$\\
  $\sigma\rv_2$ &erv2&$\kms$\\
  ${\teff}_1$ &teff1& K\\
  $\logg_1$&{logg1} & dex\\
  $\feh_{1,2}$&{feh12} &dex\\
  $\vsini_1$&{vsini1} &$\kms$\\
  ${\sigma\teff}_1$ &{eteff1} &K\\
  $\sigma\logg_1$&{elogg1} &dex\\
  $\sigma\feh_{1,2}$&{efeh12} &dex\\
  $\sigma\vsini_1$&{evsini1} &$\kms$\\
  ${\teff}_2$ &{teff2} &K\\
  $\logg_2$&{logg2} &dex\\
  $\vsini_2$&{vsini2} &$\kms$\\
  ${\sigma\teff}_2$ &{eteff2} &K\\
  $\sigma\logg_2$&{elogg2} &dex\\
  $\sigma\vsini_2$&{evsini2} &$\kms$\\
  ${\teff}_0$ &{teff0} &K\\
  $\logg_0$&{logg0} &dex\\
  $\feh_{0}$&{feh0} &dex\\
  $\vsini_0$&{vsini0} &$\kms$\\
  ${\sigma\teff}_0$ &{eteff0} &K\\
  $\sigma\logg_0$&{elogg0} &dex\\
  $\sigma\feh_{0}$&{efeh0} &dex\\
  $\sigma\vsini_0$&{evsini0} &$\kms$\\
  $\chi^2_{\rm single}$&{chi2single} &\\
  $\chi^2_{\rm binary}$&{chi2binary} &\\
  {$\imp$} &imp&\\
  $\log{k}$&{logk} &\\
  $\sigma\log{k}$&{elogk} &\\
  light fraction&{light\_fraction} &\\
  {C}&C &\\
  MJD&{mjd} & day\\
  SB2 index&{SB2} \\
  \hline
  
\end{tabular}
\label{tab:cat}
\end{table}

\section{Conclusions}
I analyzed all available spectra from the RAVE survey, using a composite spectral model. Three complementing selections found 2813 composite spectra, belonging to 2605 unique targets.
Interestingly, I found nearly the same number of SB2 candidates which was expected by \cite{rave_sb2} for a final data release of RAVE ($\sim2000$), although the selection algorithms were very different. 
 Additionally these results were compared with automatic classification based on a t-SNE map, which shows that many SB2 candidates are located in several clusters. I plan to use RVs from this study for orbit determination of SB2 candidates, which will be especially useful in addition to the future data release of Gaia data (expected on 2026/12/2). Since Gaia RVS spectra are very similar to RAVE, my selection of SB2 candidates can be easily adapted to these data, without significant modifications. Analysis of whole RAVE dataset took just a week on a personal computer with four processors, so billions of Gaia DR4 spectra can be analyzed in a reasonable time using computational cluster.  

\begin{acknowledgments}
{ I thank the anonymous referee for helpful comments and suggestions. I am also grateful to Hans Baehr (MPIA) for carefully proofreading the manuscript.}

This work is supported by the National Natural Science Foundation of China (NSFC Nos. 12288102, 12125303), the National Key R\&D Program of China (grant Nos. 2021YFA1600401, 2021YFA1600403), the Strategic Priority Research Program of the Chinese Academy of Sciences (grant Nos.  XDB1160200, XDB1160201, XDB1160300, XDB1160303, XDB1160000), the Yunnan Revitalization Talent Support Program–Science \& Technology Champion Project (No. 202305AB350003), the New Cornerstone Science Foundation through the XPLORER PRIZE, the CAS Project for Young Scientists in Basic Research (YSBR-148) and International Centre of Supernovae (ICESUN), Yunnan Key Laboratory of Supernova Research (No. 202505AV340004), the Yunnan Revitalization Talent Support Program "YunLing Scholar" project.

RAVE is an international collaboration with over 30 active members. The PI of this project is Prof. Matthias Steinmetz at Leibniz-Institute for Astrophysics Potsdam (AIP). The RAVE website and database is a service by the Leibniz-Institute for Astrophysics Potsdam (AIP). It is developed and maintained by the E-Science group.
This research has made use of NASA's Astrophysics Data System, the SIMBAD data base, and the VizieR catalogue access tool, operated at CDS, Strasbourg, France. It also made use of TOPCAT, an interactive graphical viewer and editor for tabular data \citep{topcat}. 
This work has made use of data from the European Space Agency (ESA) mission {\it Gaia} (\url{https://www.cosmos.esa.int/gaia}), processed by the {\it Gaia} Data Processing and Analysis Consortium (DPAC, \url{https://www.cosmos.esa.int/web/gaia/dpac/consortium}). Funding for the DPAC has been provided by national institutions, in particular the institutions participating in the {\it Gaia} Multilateral Agreement.

\end{acknowledgments}

%






\appendix

\section{Spectral models}
\label{sec:mod}
{ I adopted the same grid of synthetic spectra, originally computed in \cite{m11}, with the spectral systhesis code {\sc SynthV}  \citep{syntv} in LTE (local thermodynamic equilibrium) approximation and utilizing a linelist from the VALD database \citep{vald}, together with stellar atmospheric structure models, precomputed with the LL code \citep{llmod}.}

\bibliographystyle{aasjournal}

\begin{thebibliography}{}
\expandafter\ifx\csname natexlab\endcsname\relax\def\natexlab#1{#1}\fi
\providecommand{\url}[1]{\href{#1}{#1}}
\providecommand{\dodoi}[1]{doi:~\href{http://doi.org/#1}{\nolinkurl{#1}}}
\providecommand{\doeprint}[1]{\href{http://ascl.net/#1}{\nolinkurl{http://ascl.net/#1}}}
\providecommand{\doarXiv}[1]{\href{https://arxiv.org/abs/#1}{\nolinkurl{https://arxiv.org/abs/#1}}}

\bibitem[{{Birko} {et~al.}(2019){Birko}, {Zwitter}, {Grebel}, {Parker}, {Kordopatis}, {Bland-Hawthorn}, {Freeman}, {Guiglion}, {Gibson}, {Navarro}, {Reid}, {Seabroke}, {Steinmetz}, \& {Watson}}]{sb1rave2}
{Birko}, D., {Zwitter}, T., {Grebel}, E.~K., {et~al.} 2019, \aj, 158, 155, \dodoi{10.3847/1538-3881/ab3cc1}

\bibitem[{{Cropper} {et~al.}(2018){Cropper}, {Katz}, {Sartoretti}, {Prusti}, {de Bruijne}, {Chassat}, {Charvet}, {Boyadjian}, {Perryman}, {Sarri}, {Gare}, {Erdmann}, {Munari}, {Zwitter}, {Wilkinson}, {Arenou}, {Vallenari}, {G{\'o}mez}, {Panuzzo}, {Seabroke}, {Allende Prieto}, {Benson}, {Marchal}, {Huckle}, {Smith}, {Dolding}, {Jan{\ss}en}, {Viala}, {Blomme}, {Baker}, {Boudreault}, {Crifo}, {Soubiran}, {Fr{\'e}mat}, {Jasniewicz}, {Guerrier}, {Guy}, {Turon}, {Jean-Antoine-Piccolo}, {Th{\'e}venin}, {David}, {Gosset}, \& {Damerdji}}]{grvs}
{Cropper}, M., {Katz}, D., {Sartoretti}, P., {et~al.} 2018, \aap, 616, A5, \dodoi{10.1051/0004-6361/201832763}

\bibitem[{{De Silva} {et~al.}(2015){De Silva}, {Freeman}, {Bland-Hawthorn}, {Martell}, {de Boer}, {Asplund}, {Keller}, {Sharma}, {Zucker}, {Zwitter}, {Anguiano}, {Bacigalupo}, {Bayliss}, {Beavis}, {Bergemann}, {Campbell}, {Cannon}, {Carollo}, {Casagrande}, {Casey}, {Da Costa}, {D'Orazi}, {Dotter}, {Duong}, {Heger}, {Ireland}, {Kafle}, {Kos}, {Lattanzio}, {Lewis}, {Lin}, {Lind}, {Munari}, {Nataf}, {O'Toole}, {Parker}, {Reid}, {Schlesinger}, {Sheinis}, {Simpson}, {Stello}, {Ting}, {Traven}, {Watson}, {Wittenmyer}, {Yong}, \& {{\v{Z}}erjal}}]{galah2015}
{De Silva}, G.~M., {Freeman}, K.~C., {Bland-Hawthorn}, J., {et~al.} 2015, \mnras, 449, 2604, \dodoi{10.1093/mnras/stv327}

\bibitem[{{El-Badry} {et~al.}(2018{\natexlab{a}}){El-Badry}, {Rix}, {Ting}, {Weisz}, {Bergemann}, {Cargile}, {Conroy}, \& {Eilers}}]{kareem1}
{El-Badry}, K., {Rix}, H.-W., {Ting}, Y.-S., {et~al.} 2018{\natexlab{a}}, \mnras, 473, 5043, \dodoi{10.1093/mnras/stx2758}

\bibitem[{{El-Badry} {et~al.}(2018{\natexlab{b}}){El-Badry}, {Ting}, {Rix}, {Quataert}, {Weisz}, {Cargile}, {Conroy}, {Hogg}, {Bergemann}, \& {Liu}}]{bardy2018}
{El-Badry}, K., {Ting}, Y.-S., {Rix}, H.-W., {et~al.} 2018{\natexlab{b}}, \mnras, 476, 528, \dodoi{10.1093/mnras/sty240}

\bibitem[{{Gaia Collaboration} {et~al.}(2018){Gaia Collaboration}, {Brown}, {Vallenari}, {Prusti}, {de Bruijne}, {Babusiaux}, {Bailer-Jones}, {Biermann}, {Evans}, {Eyer}, \& et~al.}]{gdr2}
{Gaia Collaboration}, {Brown}, A.~G.~A., {Vallenari}, A., {et~al.} 2018, \aap, 616, A1, \dodoi{10.1051/0004-6361/201833051}

\bibitem[{{Gaia Collaboration} {et~al.}(2022){Gaia Collaboration}, {Vallenari}, {Brown}, {Prusti}, {de Bruijne}, {Arenou}, {Babusiaux}, {Biermann}, {Creevey}, {Ducourant}, {Evans}, {Eyer}, {Guerra}, {Hutton}, {Jordi}, {Klioner}, {Lammers}, {Lindegren}, {Luri}, {Mignard}, {Panem}, {Pourbaix}, {Randich}, {Sartoretti}, {Soubiran}, {Tanga}, {Walton}, {Bailer-Jones}, {Bastian}, {Drimmel}, {Jansen}, {Katz}, {Lattanzi}, {van Leeuwen}, {Bakker}, {Cacciari}, {Casta{\~n}eda}, {De Angeli}, {Fabricius}, {Fouesneau}, {Fr{\'e}mat}, {Galluccio}, {Guerrier}, {Heiter}, {Masana}, {Messineo}, {Mowlavi}, {Nicolas}, {Nienartowicz}, {Pailler}, {Panuzzo}, {Riclet}, {Roux}, {Seabroke}, {Sordo{\o}rcit}, {Th{\'e}venin}, {Gracia-Abril}, {Portell}, {Teyssier}, {Altmann}, {Andrae}, {Audard}, {Bellas-Velidis}, {Benson}, {Berthier}, {Blomme}, {Burgess}, {Busonero}, {Busso}, {C{\'a}novas}, {Carry}, {Cellino}, {Cheek}, {Clementini}, {Damerdji}, {Davidson}, {de Teodoro}, {Nu{\~n}ez Campos}, {Delchambre}, {Dell'Oro}, {Esquej},
  {Fern{\'a}ndez-Hern{\'a}ndez}, {Fraile}, {Garabato}, {Garc{\'\i}a-Lario}, {Gosset}, {Haigron}, {Halbwachs}, {Hambly}, {Harrison}, {Hern{\'a}ndez}, {Hestroffer}, {Hodgkin}, {Holl}, {Jan{\ss}en}, {Jevardat de Fombelle}, {Jordan}, {Krone-Martins}, {Lanzafame}, {L{\"o}ffler}, {Marchal}, {Marrese}, {Moitinho}, {Muinonen}, {Osborne}, {Pancino}, {Pauwels}, {Recio-Blanco}, {Reyl{\'e}}, {Riello}, {Rimoldini}, {Roegiers}, {Rybizki}, {Sarro}, {Siopis}, {Smith}, {Sozzetti}, {Utrilla}, {van Leeuwen}, {Abbas}, {{\'A}brah{\'a}m}, {Abreu Aramburu}, {Aerts}, {Aguado}, {Ajaj}, {Aldea-Montero}, {Altavilla}, {{\'A}lvarez}, {Alves}, {Anders}, {Anderson}, {Anglada Varela}, {Antoja}, {Baines}, {Baker}, {Balaguer-N{\'u}{\~n}ez}, {Balbinot}, {Balog}, {Barache}, {Barbato}, {Barros}, {Barstow}, {Bartolom{\'e}}, {Bassilana}, {Bauchet}, {Becciani}, {Bellazzini}, {Berihuete}, {Bernet}, {Bertone}, {Bianchi}, {Binnenfeld}, {Blanco-Cuaresma}, {Blazere}, {Boch}, {Bombrun}, {Bossini}, {Bouquillon}, {Bragaglia}, {Bramante}, {Breedt},
  {Bressan}, {Brouillet}, {Brugaletta}, {Bucciarelli}, {Burlacu}, {Butkevich}, {Buzzi}, {Caffau}, {Cancelliere}, {Cantat-Gaudin}, {Carballo}, {Carlucci}, {Carnerero}, {Carrasco}, {Casamiquela}, {Castellani}, {Castro-Ginard}, {Chaoul}, {Charlot}, {Chemin}, {Chiaramida}, {Chiavassa}, {Chornay}, {Comoretto}, {Contursi}, {Cooper}, {Cornez}, {Cowell}, {Crifo}, {Cropper}, {Crosta}, {Crowley}, {Dafonte}, {Dapergolas}, {David}, {David}, {de Laverny}, {De Luise}, {De March}, {De Ridder}, {de Souza}, {de Torres}, {del Peloso}, {del Pozo}, {Delbo}, {Delgado}, {Delisle}, {Demouchy}, {Dharmawardena}, {Di Matteo}, {Diakite}, {Diener}, {Distefano}, {Dolding}, {Edvardsson}, {Enke}, {Fabre}, {Fabrizio}, {Faigler}, {Fedorets}, {Fernique}, {Fienga}, {Figueras}, {Fournier}, {Fouron}, {Fragkoudi}, {Gai}, {Garcia-Gutierrez}, {Garcia-Reinaldos}, {Garc{\'\i}a-Torres}, {Garofalo}, {Gavel}, {Gavras}, {Gerlach}, {Geyer}, {Giacobbe}, {Gilmore}, {Girona}, {Giuffrida}, {Gomel}, {Gomez}, {Gonz{\'a}lez-N{\'u}{\~n}ez},
  {Gonz{\'a}lez-Santamar{\'\i}a}, {Gonz{\'a}lez-Vidal}, {Granvik}, {Guillout}, {Guiraud}, {Guti{\'e}rrez-S{\'a}nchez}, {Guy}, {Hatzidimitriou}, {Hauser}, {Haywood}, {Helmer}, {Helmi}, {Sarmiento}, {Hidalgo}, {Hilger}, {H{\l}adczuk}, {Hobbs}, {Holland}, {Huckle}, {Jardine}, {Jasniewicz}, {Jean-Antoine Piccolo}, {Jim{\'e}nez-Arranz}, {Jorissen}, {Juaristi Campillo}, {Julbe}, {Karbevska}, {Kervella}, {Khanna}, {Kontizas}, {Kordopatis}, {Korn}, {K{\'o}sp{\'a}l}, {Kostrzewa-Rutkowska}, {Kruszy{\'n}ska}, {Kun}, {Laizeau}, {Lambert}, {Lanza}, {Lasne}, {Le Campion}, {Lebreton}, {Lebzelter}, {Leccia}, {Leclerc}, {Lecoeur-Taibi}, {Liao}, {Licata}, {Lindstr{\o}m}, {Lister}, {Livanou}, {Lobel}, {Lorca}, {Loup}, {Madrero Pardo}, {Magdaleno Romeo}, {Managau}, {Mann}, {Manteiga}, {Marchant}, {Marconi}, {Marcos}, {Marcos Santos}, {Mar{\'\i}n Pina}, {Marinoni}, {Marocco}, {Marshall}, {Polo}, {Mart{\'\i}n-Fleitas}, {Marton}, {Mary}, {Masip}, {Massari}, {Mastrobuono-Battisti}, {Mazeh}, {McMillan}, {Messina}, {Michalik},
  {Millar}, {Mints}, {Molina}, {Molinaro}, {Moln{\'a}r}, {Monari}, {Mongui{\'o}}, {Montegriffo}, {Montero}, {Mor}, {Mora}, {Morbidelli}, {Morel}, {Morris}, {Muraveva}, {Murphy}, {Musella}, {Nagy}, {Noval}, {Oca{\~n}a}, {Ogden}, {Ordenovic}, {Osinde}, {Pagani}, {Pagano}, {Palaversa}, {Palicio}, {Pallas-Quintela}, {Panahi}, {Payne-Wardenaar}, {Pe{\~n}alosa Esteller}, {Penttil{\"a}}, {Pichon}, {Piersimoni}, {Pineau}, {Plachy}, {Plum}, {Poggio}, {Pr{\v{s}}a}, {Pulone}, {Racero}, {Ragaini}, {Rainer}, {Raiteri}, {Rambaux}, {Ramos}, {Ramos-Lerate}, {Re Fiorentin}, {Regibo}, {Richards}, {Rios Diaz}, {Ripepi}, {Riva}, {Rix}, {Rixon}, {Robichon}, {Robin}, {Robin}, {Roelens}, {Rogues}, {Rohrbasser}, {Romero-G{\'o}mez}, {Rowell}, {Royer}, {Ruz Mieres}, {Rybicki}, {Sadowski}, {S{\'a}ez N{\'u}{\~n}ez}, {Sagrist{\`a} Sell{\'e}s}, {Sahlmann}, {Salguero}, {Samaras}, {Sanchez Gimenez}, {Sanna}, {Santove{\~n}a}, {Sarasso}, {Schultheis}, {Sciacca}, {Segol}, {Segovia}, {S{\'e}gransan}, {Semeux}, {Shahaf}, {Siddiqui}, {Siebert},
  {Siltala}, {Silvelo}, {Slezak}, {Slezak}, {Smart}, {Snaith}, {Solano}, {Solitro}, {Souami}, {Souchay}, {Spagna}, {Spina}, {Spoto}, {Steele}, {Steidelm{\"u}ller}, {Stephenson}, {S{\"u}veges}, {Surdej}, {Szabados}, {Szegedi-Elek}, {Taris}, {Taylo}, {Teixeira}, {Tolomei}, {Tonello}, {Torra}, {Torra}, {Torralba Elipe}, {Trabucchi}, {Tsounis}, {Turon}, {Ulla}, {Unger}, {Vaillant}, {van Dillen}, {van Reeven}, {Vanel}, {Vecchiato}, {Viala}, {Vicente}, {Voutsinas}, {Weiler}, {Wevers}, {Wyrzykowski}, {Yoldas}, {Yvard}, {Zhao}, {Zorec}, {Zucker}, \& {Zwitter}}]{gaia3}
{Gaia Collaboration}, {Vallenari}, A., {Brown}, A.~G.~A., {et~al.} 2022, arXiv e-prints, arXiv:2208.00211.
\newblock \doarXiv{2208.00211}

\bibitem[{{Gilmore} {et~al.}(2012){Gilmore}, {Randich}, {Asplund}, {Binney}, {Bonifacio}, {Drew}, {Feltzing}, {Ferguson}, {Jeffries}, {Micela}, {Negueruela}, {Prusti}, {Rix}, {Vallenari}, {Alfaro}, {Allende-Prieto}, {Babusiaux}, {Bensby}, {Blomme}, {Bragaglia}, {Flaccomio}, {Fran{\c{c}}ois}, {Irwin}, {Koposov}, {Korn}, {Lanzafame}, {Pancino}, {Paunzen}, {Recio-Blanco}, {Sacco}, {Smiljanic}, {Van Eck}, {Walton}, {Aden}, {Aerts}, {Affer}, {Alcala}, {Altavilla}, {Alves}, {Antoja}, {Arenou}, {Argiroffi}, {Asensio Ramos}, {Bailer-Jones}, {Balaguer-Nunez}, {Bayo}, {Barbuy}, {Barisevicius}, {Barrado y Navascues}, {Battistini}, {Bellas Velidis}, {Bellazzini}, {Belokurov}, {Bergemann}, {Bertelli}, {Biazzo}, {Bienayme}, {Bland-Hawthorn}, {Boeche}, {Bonito}, {Boudreault}, {Bouvier}, {Brandao}, {Brown}, {de Bruijne}, {Burleigh}, {Caballero}, {Caffau}, {Calura}, {Capuzzo-Dolcetta}, {Caramazza}, {Carraro}, {Casagrande}, {Casewell}, {Chapman}, {Chiappini}, {Chorniy}, {Christlieb}, {Cignoni}, {Cocozza}, {Colless}, {Collet},
  {Collins}, {Correnti}, {Covino}, {Crnojevic}, {Cropper}, {Cunha}, {Damiani}, {David}, {Delgado}, {Duffau}, {Edvardsson}, {Eldridge}, {Enke}, {Eriksson}, {Evans}, {Eyer}, {Famaey}, {Fellhauer}, {Ferreras}, {Figueras}, {Fiorentino}, {Flynn}, {Folha}, {Franciosini}, {Frasca}, {Freeman}, {Fremat}, {Friel}, {Gaensicke}, {Gameiro}, {Garzon}, {Geier}, {Geisler}, {Gerhard}, {Gibson}, {Gomboc}, {Gomez}, {Gonzalez-Fernandez}, {Gonzalez Hernandez}, {Gosset}, {Grebel}, {Greimel}, {Groenewegen}, {Grundahl}, {Guarcello}, {Gustafsson}, {Hadrava}, {Hatzidimitriou}, {Hambly}, {Hammersley}, {Hansen}, {Haywood}, {Heber}, {Heiter}, {Held}, {Helmi}, {Hensler}, {Herrero}, {Hill}, {Hodgkin}, {Huelamo}, {Huxor}, {Ibata}, {Jackson}, {de Jong}, {Jonker}, {Jordan}, {Jordi}, {Jorissen}, {Katz}, {Kawata}, {Keller}, {Kharchenko}, {Klement}, {Klutsch}, {Knude}, {Koch}, {Kochukhov}, {Kontizas}, {Koubsky}, {Lallement}, {de Laverny}, {van Leeuwen}, {Lemasle}, {Lewis}, {Lind}, {Lindstrom}, {Lobel}, {Lopez Santiago}, {Lucas}, {Ludwig},
  {Lueftinger}, {Magrini}, {Maiz Apellaniz}, {Maldonado}, {Marconi}, {Marino}, {Martayan}, {Martinez-Valpuesta}, {Matijevic}, {McMahon}, {Messina}, {Meyer}, {Miglio}, {Mikolaitis}, {Minchev}, {Minniti}, {Moitinho}, {Momany}, {Monaco}, {Montalto}, {Monteiro}, {Monier}, {Montes}, {Mora}, {Moraux}, {Morel}, \& {Mowlavi}}]{ges}
{Gilmore}, G., {Randich}, S., {Asplund}, M., {et~al.} 2012, The Messenger, 147, 25

\bibitem[{{Guo} {et~al.}(2025){Guo}, {Kovalev}, {Li}, {L{\"u}}, {Jia}, {Li}, {Li}, {Xiong}, {Yang}, {He}, {Chen}, \& {Han}}]{guo665}
{Guo}, S., {Kovalev}, M., {Li}, J., {et~al.} 2025, \apjs, 278, 46, \dodoi{10.3847/1538-4365/adced1}

\bibitem[{{Jack}(2019)}]{sb2jack_2019AN....340..386J}
{Jack}, D. 2019, Astronomische Nachrichten, 340, 386, \dodoi{10.1002/asna.201913496}

\bibitem[{{Jayasinghe} {et~al.}(2018){Jayasinghe}, {Kochanek}, {Stanek}, {Shappee}, {Holoien}, {Thompson}, {Prieto}, {Dong}, {Pawlak}, {Shields}, {Pojmanski}, {Otero}, {Britt}, \& {Will}}]{asassnv}
{Jayasinghe}, T., {Kochanek}, C.~S., {Stanek}, K.~Z., {et~al.} 2018, \mnras, 477, 3145, \dodoi{10.1093/mnras/sty838}

\bibitem[{{Kounkel} {et~al.}(2021){Kounkel}, {Covey}, {Stassun}, {Price-Whelan}, {Holtzman}, {Chojnowski}, {Longa-Pe{\~n}a}, {Rom{\'a}n-Z{\'u}{\~n}iga}, {Hernandez}, {Serna}, {Badenes}, {De Lee}, {Majewski}, {Stringfellow}, {Kratter}, {Moe}, {Frinchaboy}, {Beaton}, {Fern{\'a}ndez-Trincado}, {Mahadevan}, {Minniti}, {Beers}, {Schneider}, {Barba}, {Brownstein}, {Garc{\'\i}a-Hern{\'a}ndez}, {Pan}, \& {Bizyaev}}]{kounkel}
{Kounkel}, M., {Covey}, K.~R., {Stassun}, K.~G., {et~al.} 2021, \aj, 162, 184, \dodoi{10.3847/1538-3881/ac1798}

\bibitem[{{Kovalev} {et~al.}(2024{\natexlab{a}}){Kovalev}, {Ahmed}, \& {Asa'd}}]{rot5oc}
{Kovalev}, M., {Ahmed}, M., \& {Asa'd}, R. 2024{\natexlab{a}}, \mnras, 527, 9595, \dodoi{10.1093/mnras/stad3833}

\bibitem[{{Kovalev} {et~al.}(2022){Kovalev}, {Chen}, \& {Han}}]{cat22}
{Kovalev}, M., {Chen}, X., \& {Han}, Z. 2022, \mnras, 517, 356, \dodoi{10.1093/mnras/stac2513}

\bibitem[{{Kovalev} \& {Straumit}(2022)}]{m11}
{Kovalev}, M., \& {Straumit}, I. 2022, \mnras, 510, 1515, \dodoi{10.1093/mnras/stab3365}

\bibitem[{{Kovalev} {et~al.}(2024{\natexlab{b}}){Kovalev}, {Zhou}, {Chen}, \& {Han}}]{cat23}
{Kovalev}, M., {Zhou}, Z., {Chen}, X., \& {Han}, Z. 2024{\natexlab{b}}, \mnras, 527, 521, \dodoi{10.1093/mnras/stad3222}

\bibitem[{{Kovalev} {et~al.}(2024{\natexlab{c}}){Kovalev}, {Chen}, \& {Han}}]{j115}
{Kovalev}, M.~Y., {Chen}, X., \& {Han}, Z. 2024{\natexlab{c}}, Research Notes of the American Astronomical Society, 8, 175, \dodoi{10.3847/2515-5172/ad5f2f}

\bibitem[{Kovalev {et~al.}(2026)Kovalev, Kniazev, \& Malkov}]{galaxies14020027}
Kovalev, M.~Y., Kniazev, A.~Y., \& Malkov, O.~Y. 2026, Galaxies, 14, \dodoi{10.3390/galaxies14020027}

\bibitem[{{Liu} {et~al.}(2020){Liu}, {Fu}, {Shi}, {Wu}, {Han}, {Chen}, {Dong}, {Zhao}, {Chen}, {Zhang}, {Bai}, {Chen}, {Cui}, {Du}, {Hsia}, {Jiang}, {Hou}, {Hou}, {Li}, {Li}, {Li}, {Liu}, {Liu}, {Luo}, {Ren}, {Tian}, {Tian}, {Wang}, {Wu}, {Xie}, {Yan}, {Yang}, {Yu}, {Zhang}, {Zhang}, {Zhang}, {Zhang}, {Zhao}, {Zhong}, {Zong}, \& {Zuo}}]{mrs}
{Liu}, C., {Fu}, J., {Shi}, J., {et~al.} 2020, arXiv e-prints, arXiv:2005.07210.
\newblock \doarXiv{2005.07210}

\bibitem[{{Matijevi{\v{c}}} {et~al.}(2010){Matijevi{\v{c}}}, {Zwitter}, {Munari}, {Bienaym{\'e}}, {Binney}, {Bland-Hawthorn}, {Boeche}, {Campbell}, {Freeman}, {Gibson}, {Gilmore}, {Grebel}, {Helmi}, {Navarro}, {Parker}, {Seabroke}, {Siebert}, {Siviero}, {Steinmetz}, {Watson}, {Williams}, \& {Wyse}}]{rave_sb2}
{Matijevi{\v{c}}}, G., {Zwitter}, T., {Munari}, U., {et~al.} 2010, \aj, 140, 184, \dodoi{10.1088/0004-6256/140/1/184}

\bibitem[{{Matijevi{\v{c}}} {et~al.}(2011){Matijevi{\v{c}}}, {Zwitter}, {Bienaym{\'e}}, {Bland-Hawthorn}, {Freeman}, {Gilmore}, {Grebel}, {Helmi}, {Munari}, {Navarro}, {Parker}, {Reid}, {Seabroke}, {Siebert}, {Siviero}, {Steinmetz}, {Watson}, {Williams}, \& {Wyse}}]{rave_sb1}
{Matijevi{\v{c}}}, G., {Zwitter}, T., {Bienaym{\'e}}, O., {et~al.} 2011, \aj, 141, 200, \dodoi{10.1088/0004-6256/141/6/200}

\bibitem[{{Matijevi{\v{c}}} {et~al.}(2012){Matijevi{\v{c}}}, {Zwitter}, {Bienaym{\'e}}, {Bland-Hawthorn}, {Boeche}, {Freeman}, {Gibson}, {Gilmore}, {Grebel}, {Helmi}, {Munari}, {Navarro}, {Parker}, {Reid}, {Seabroke}, {Siebert}, {Siviero}, {Steinmetz}, {Watson}, {Williams}, \& {Wyse}}]{ravelle}
---. 2012, \apjs, 200, 14, \dodoi{10.1088/0067-0049/200/2/14}

\bibitem[{{Pourbaix} {et~al.}(2004){Pourbaix}, {Tokovinin}, {Batten}, {Fekel}, {Hartkopf}, {Levato}, {Morrell}, {Torres}, \& {Udry}}]{sb9}
{Pourbaix}, D., {Tokovinin}, A.~A., {Batten}, A.~H., {et~al.} 2004, \aap, 424, 727, \dodoi{10.1051/0004-6361:20041213}

\bibitem[{{Rowan} {et~al.}(2022){Rowan}, {Jayasinghe}, {Stanek}, {Kochanek}, {Thompson}, {Shappee}, {Holoien}, {Prieto}, \& {Giles}}]{rowan_asassin}
{Rowan}, D.~M., {Jayasinghe}, T., {Stanek}, K.~Z., {et~al.} 2022, \mnras, 517, 2190, \dodoi{10.1093/mnras/stac2520}

\bibitem[{{Ryabchikova} {et~al.}(2015){Ryabchikova}, {Piskunov}, {Kurucz}, {Stempels}, {Heiter}, {Pakhomov}, \& {Barklem}}]{vald}
{Ryabchikova}, T., {Piskunov}, N., {Kurucz}, R.~L., {et~al.} 2015, \physscr, 90, 054005, \dodoi{10.1088/0031-8949/90/5/054005}

\bibitem[{{Shulyak} {et~al.}(2004){Shulyak}, {Tsymbal}, {Ryabchikova}, {St{\"u}tz}, \& {Weiss}}]{llmod}
{Shulyak}, D., {Tsymbal}, V., {Ryabchikova}, T., {St{\"u}tz}, C., \& {Weiss}, W.~W. 2004, \aap, 428, 993, \dodoi{10.1051/0004-6361:20034169}

\bibitem[{{Steinmetz} {et~al.}(2018){Steinmetz}, {Zwitter}, {Matijevic}, {Siviero}, \& {Munari}}]{problem_rv5}
{Steinmetz}, M., {Zwitter}, T., {Matijevic}, G., {Siviero}, A., \& {Munari}, U. 2018, Research Notes of the American Astronomical Society, 2, 194, \dodoi{10.3847/2515-5172/aaead0}

\bibitem[{{Steinmetz} {et~al.}(2020){Steinmetz}, {Matijevi{\v{c}}}, {Enke}, {Zwitter}, {Guiglion}, {McMillan}, {Kordopatis}, {Valentini}, {Chiappini}, {Casagrande}, {Wojno}, {Anguiano}, {Bienaym{\'e}}, {Bijaoui}, {Binney}, {Burton}, {Cass}, {de Laverny}, {Fiegert}, {Freeman}, {Fulbright}, {Gibson}, {Gilmore}, {Grebel}, {Helmi}, {Kunder}, {Munari}, {Navarro}, {Parker}, {Ruchti}, {Recio-Blanco}, {Reid}, {Seabroke}, {Siviero}, {Siebert}, {Stupar}, {Watson}, {Williams}, {Wyse}, {Anders}, {Antoja}, {Birko}, {Bland-Hawthorn}, {Bossini}, {Garc{\'\i}a}, {Carrillo}, {Chaplin}, {Elsworth}, {Famaey}, {Gerhard}, {Jofre}, {Just}, {Mathur}, {Miglio}, {Minchev}, {Monari}, {Mosser}, {Ritter}, {Rodrigues}, {Scholz}, {Sharma}, {Sysoliatina}, \& {RAVE Collaboration}}]{rave6}
{Steinmetz}, M., {Matijevi{\v{c}}}, G., {Enke}, H., {et~al.} 2020, \aj, 160, 82, \dodoi{10.3847/1538-3881/ab9ab9}

\bibitem[{{Sysoliatina} {et~al.}(2018){Sysoliatina}, {Just}, {Koutsouridou}, {Grebel}, {Kordopatis}, {Steinmetz}, {Bienaym{\'e}}, {Gibson}, {Navarro}, {Reid}, \& {Seabroke}}]{xenia}
{Sysoliatina}, K., {Just}, A., {Koutsouridou}, I., {et~al.} 2018, \aap, 620, A71, \dodoi{10.1051/0004-6361/201833228}

\bibitem[{{Tanikawa} {et~al.}(2026){Tanikawa}, {Tajitsu}, {Honda}, {Maehara}, {Sato}, {Masuda}, {Omiya}, \& {Izumiura}}]{sb3japan}
{Tanikawa}, A., {Tajitsu}, A., {Honda}, S., {et~al.} 2026, arXiv e-prints, arXiv:2601.21125, \dodoi{10.48550/arXiv.2601.21125}

\bibitem[{{Taylor}(2005)}]{topcat}
{Taylor}, M.~B. 2005, in Astronomical Society of the Pacific Conference Series, Vol. 347, Astronomical Data Analysis Software and Systems XIV, ed. P.~{Shopbell}, M.~{Britton}, \& R.~{Ebert}, 29

\bibitem[{{Traven} {et~al.}(2017){Traven}, {Matijevi{\v{c}}}, {Zwitter}, {{\v{Z}}erjal}, {Kos}, {Asplund}, {Bland-Hawthorn}, {Casey}, {De Silva}, {Freeman}, {Lin}, {Martell}, {Schlesinger}, {Sharma}, {Simpson}, {Zucker}, {Anguiano}, {Da Costa}, {Duong}, {Horner}, {Hyde}, {Kafle}, {Munari}, {Nataf}, {Navin}, {Reid}, \& {Ting}}]{traven_tsne}
{Traven}, G., {Matijevi{\v{c}}}, G., {Zwitter}, T., {et~al.} 2017, \apjs, 228, 24, \dodoi{10.3847/1538-4365/228/2/24}

\bibitem[{{Tsymbal}(1996)}]{syntv}
{Tsymbal}, V. 1996, in Astronomical Society of the Pacific Conference Series, Vol. 108, M.A.S.S., Model Atmospheres and Spectrum Synthesis, ed. S.~J. {Adelman}, F.~{Kupka}, \& W.~W. {Weiss}, 198

\bibitem[{Van Der~Maaten \& Hinton(2008)}]{tsne}
Van Der~Maaten, L. J.~P., \& Hinton, G.~E. 2008, 9, 2579, \dodoi{10.1007/s10479-011-0841-3}

\end{thebibliography}


\end{document}